\documentclass[12pt]{article}
\usepackage[utf8]{inputenc}
\usepackage{graphicx} 
\usepackage{dcolumn} 
\usepackage{bm} 
\usepackage{amsfonts}
\usepackage{amsthm}
\usepackage{amsmath}
\usepackage{amssymb}
\usepackage{epsfig}
\usepackage{color}
\usepackage{textcomp}
\usepackage{hyperref}
\usepackage{csquotes}
\usepackage{titlesec}
\usepackage{slashed}
\usepackage{caption}
\usepackage{subcaption}
\usepackage{booktabs}
\usepackage{multirow}
\usepackage{cite}
\usepackage{placeins}
\usepackage{comment}
\usepackage{braket}
\usepackage{authblk}
\usepackage{physics}
\usepackage{mathtools}
\numberwithin{equation}{section}

\title{\textbf{Time as a Cosmological Phenomenon}}
\author{Andrea Palessandro \thanks{andrea.palessandro@gmail.com}}
\affil{\small Società Italiana di Fisica}
\date{}

\begin{document}

\maketitle
\begin{abstract}
    \noindent We show that the arrow of time is intimately related to the geometry and topology of the whole universe, and is therefore best understood as a cosmological phenomenon.
\end{abstract}

\noindent\rule{\linewidth}{0.4pt}

{\small
\begin{quote}
Non è suo moto per altro distinto,\\
ma li altri son mensurati da questo,\\
sì come diece da mezzo e da quinto;\\

e come il tempo tegna in cotal testo\\
le sue radici e ne li altri le fronde,\\
omai a te può esser manifesto.
\begin{flushright}
\textit{Paradiso, Canto XXVII}
\end{flushright}
\end{quote}
}

\noindent\rule{\linewidth}{0.4pt}

\section{Introduction}
\noindent There are two opposing views in the philosophy of time, sometimes called the A theory and the B theory of time \cite{McTaggart}. Roughly speaking, the A theory asserts that the flow of time is real, and, as a consequence, the future does not yet exist\footnote{Depending on whether the \textit{past} is deemed to still exist or not, the A theory further divides into the \textit{growing block} view and the \textit{presentist} view, respectively.}, while the B theory maintains that time is an illusion, and past, present and future are equally real\footnote{This is sometimes called \textit{eternalism}, or the block universe theory, according to which temporal becoming is not an objective feature of reality, but a subjective illusion of human consciousness.} \cite{time}. The two views reflect an ancient philosophical debate that goes back at least to Heraclitus (advocate of the A theory) and Parmenides (advocate of the B theory).

There are two ways to define time in physics that roughly correspond to the A and B theory ontologies. In classical physics\footnote{By classical physics I here mean non-quantum physics, so that includes the general theory of relativity.}, time is essentially defined as a coordinate that identifies points in spacetime: in the theory of relativity, spacetime is a four-dimensional manifold parametrized by four numbers, one of which is time. On this manifold, all points are equally real, therefore it is impossible to define a passage of time, as in the B theory.

The chief argument that supports this conclusion was developed independently by Rietdijk, Putnam and Penrose \cite{Rietdijk, Putnam, Penrose1}. They argue that, according to (special) relativity, different observers moving at different velocities with respect to each other will have different planes of simultaneity, and hence different sets of events that constitute their present moment. If one takes the view that only the present exists (or at most the present and the past), the three-dimensional slice of the universe that is considered to exist at any given moment will vary depending on the observer, which violates a fundamental tenet of relativity theory.

Even if it was somehow possible to define a universal ``present moment'', as in Newtonian physics, it would still be impossible to define a direction, or arrow, of time. This is because all known fundamental laws of physics are time-reversible and cannot really distinguish between forward and backward time evolution. So while it might be possible in Newtonian physics to maintain that only the present moment exists, as its content is the same for all observers, it would still be impossible to determine in which direction it is evolving. 

While unambiguous, then, the B-theory definition of time in physics is unsatisfactory insofar as it does not conform with the human experience of time and memory. In General Relativity, time serves merely as a useful parametrization of events, and has no relation with what we commonly experience as ``time'' as a constant stream of changing events. The time of experience is instead captured much more closely by ``thermodynamic'' time, that evolves by definition in the direction of increasing entropy. This is the A-theory definition of time in physics\footnote{While thermodynamic time is compatible with the A-theory concept of time as ``flow'', it is a matter of dispute whether this means that the future does not yet exist in a classical thermodynamic system. It has been argued that this can only be cogently said of quantum systems, in which there exist a fundamental difference between the certain past and the uncertain future via non-deterministic wave function collapse, see \cite{Ellis1, Ellis2, Ellis3}.}.

Indeed, while there is no way to consistently define an arrow of time from microphysics alone (due to the relativity of simultaneity and the time-reversibility of the dynamical laws), its existence is certainly compatible with, albeit not mandated by, macrophysics, particularly thermodynamics. Whenever the spacetime manifold is entropically asymmetric, in the sense that it includes a relatively ordered low-entropy region\footnote{In our own universe, this corresponds to the Big Bang.} together with a relatively disordered high-entropy region, it is always possible to define, at least locally, an arrow of time that connects the two, going from order to disorder \cite{Price}. The second law of thermodynamics gives a simple procedure to distinguish the past from the future \cite{Penrose3} and is ultimately what gives rise to memory, and consequently to the psychological experience of the passage of time \cite{Wolpert}. 

However, and crucially, General Relativity is not generally compatible with the second law of thermodynamics, meaning that there are solutions of Einstein's equations for which no consistent formulation of the second law can be given. In other words, while the time of microphysics (the B-theory definition of time as coordinate of spacetime events) is always well-defined, the time of macrophysics (the A-theory definition of time as entropy increase), as we will see, is not. 

The fact that entropy can and does increase in our universe is not a fundamental law, and can only be understood as a cosmological fact, namely as the boundary conditions that gave rise to a time-orientable, synchronous, chronological, geodesically incomplete universe with vanishing Weyl tensor at the initial singularity. Were even one of the above assumptions false, for example in a universe that is not time-orientable \cite{Hadley}, a universe with closed timelike curves \cite{Godel}, or a universe initially in gravitational equilibrium \cite{Penrose2}, no coherent notion of time (and its passage) could be formulated.

In light of these considerations, Earman \cite{Earman} was the first to suggest that the arrow of time might be a geometrical feature of the universe. Rather than grounding temporal asymmetry in thermodynamic boundary conditions, he proposed that directionality could be encoded in global spacetime structure, causal ordering, and related topological properties. According to Earman's \textit{time direction heresy}, ``temporal orientation is an intrinsic feature of space-time which does not need to be and cannot be reduced to nontemporal features''. This view was later expanded by Castagnino, Lombardi, and others in a series of papers \cite{Castagnino1, Castagnino2, Castagnino3, Castagnino4, Castagnino5, Castagnino6, Castagnino7, Castagnino8}.

Our view differs from Earman's heresy (while being indebted to it) in that we maintain that nontemporal features, such as entropy, are indeed essential for defining a temporal orientation. However, these nontemporal features can themselves be meaningfully specified only when the underlying spacetime possesses the appropriate geometrical and topological properties. Accordingly, we assign primacy, but not exclusivity, to the geometrical and topological structure: it forms the necessary framework within which nontemporal asymmetries can be properly formulated and interpreted.

In the following sections we give several examples of universes (cosmological solutions of Einstein's equations) that admit no macroscopic arrow of time. This could be for two reasons:
\begin{itemize}
    \item \textbf{Local obstructions}: The universe does not admit a well defined notion of cosmic time, or, if it does, it started in a high-entropy state, i.e. either Weyl's postulate \cite{Rugh2} or the past hypothesis \cite{Ainsworth} (or both) are violated. Geometrical obstructions of this type prevent even a local arrow of time from being defined.
    \item \textbf{Global obstructions}: The universe is not time-orientable \cite{Lemos} or admits closed timelike curves (CTC) \cite{Thorne} that violate causality and along which entropy cannot globally increase, i.e. the universe is not chronological \cite{Hawking}. Causal or topological obstructions of this type prevent a global arrow of time from being defined everywhere on the spacetime manifold.
\end{itemize}

There could be spacetimes that admit a local but not global arrow of time \cite{Matthews, Rugh, Ellis4}\footnote{Ellis in particular says about the arrow of time locality issue that ``local determination has to arbitrarily choose one of the two directions of time as the positive direction indicating the future; but as this decision is made locally, there is no reason whatever why it should be consistent globally. If it emerges locally, opposite arrows may be expected to occur in different places''.}, for example spacetimes that satisfy Weyl's postulate and the past hypothesis, but are nevertheless not chronological due to a non-trivial spacetime topology which prevents the local arrow of time from being extended globally. In this paper, we analyze the conditions that must be true for the arrow of time to exist globally. In order, these are
\begin{enumerate}
    \item Weyl's postulate, which we analyze in \S \ref{Weyl}.
    \item The past hypothesis, which we analyze in \S \ref{past}.
    \item Time-orientability, which we analyze in \S \ref{Mobius}.
    \item Chronology, which we analyze in \S \ref{chronology}.
\end{enumerate}
Each condition builds upon the previous and places further constraints on the shape the universe needs to have in order to allow for the experience of time. If any of the conditions above is violated, the universe will not have a measurable arrow of time (at least globally), and existence will quite literally be timeless\footnote{There might very well be other conditions beyond the three we have listed above that are necessary for the existence of experiential time, but none of them will alone be sufficient to guarantee it.}.

In \S \ref{entropy} we discuss in more details the role of entropy in defining the arrow of time in a cosmological setting, and compare and contrast our approach with that of Castagnino and Lombardi. Finally, in \S \ref{POT} we discuss the quantum perspective on this issue, and present our conclusions in \S \ref{conclusions}.

\section{Weyl's Postulate}\label{Weyl}
Given a spacetime manifold $\mathcal{M}$ with metric $g$ and a local coordinate chart $x^\mu = (t, \vec{x})$\footnote{We use Greek letters for spacetime indices and Latin letters for space indices.}, it is in general not possible to define on it a direction of time, even locally. This is for essentially two reasons. 

First, it is not a priori clear \textit{whose} time $t$ should be chosen. In general, different observers measure time differently and have different planes of simultaneity, so that there is no universal notion of ``the passage of time''. Another more technical way of saying this is that there is no universally agreed upon \textit{foliation} of spacetime in the general case: different observers will have different planes of simultaneity, which in turn correspond to different time directions (orthogonal to the plane). There is no objective criterion that allows one to select a preferred spacelike hypersurface, and equivalently a preferred time direction, for a general spacetime metric.

Second, Einstein's equations are invariant under time-reversal, i.e. acting on a solution with the transformation $t \rightarrow -t$ gives back another perfectly valid solution. Ergo, even when a universal cosmic time direction can be defined, it is not possible from the structure of the equations alone to determine its orientation (sometimes also called ``sense''). Something else is needed, and this something has to do with the initial distribution of the matter content in the universe, or, equivalently, with its initial geometry, as we will see\footnote{In the next section we show that in order for time to have a well-defined orientation, the universe has to be globally geodesically incomplete, meaning that no geodesic belonging to the congruence defined in this section can be indefinitely extended in one of the two time directions. This gives a well-defined notion of ``initial'', which can be taken to mean ``starting from the initial singularity''.}.

An ``arrow of time'' will only emerge when we have solved both problems above, i.e. when we have managed to give time, via suitable initial conditions, both a \textit{direction} and an \textit{orientation}.

Let's start with the first problem, namely that of giving time a universal direction. The most general metric we can put on our spacetime manifold $\mathcal{M}$ is
\begin{equation}\label{metric}
    ds^2 = g_{\mu \nu}(t,\vec{x}) dx^\mu dx^\nu.
\end{equation}
The existence of a universal cosmic time is contingent upon the existence of a preferred spatial slicing on $\mathcal{M}$. Given that the metric $g$ is uniquely determined by the matter distribution on $\mathcal{M}$, a preferred spatial slicing can only be defined using the world lines of the fluid particles that live on the manifold. This is exactly the strategy adopted by Weyl's postulate \cite{Narlikar}:

\begin{quote}
\textbf{Weyl's Postulate}\\
The preferred spatial slicing is given by taking the unique hypersurfaces orthogonal to the world lines of the fluid particles.
\end{quote}

\begin{figure}
  \centering
  \includegraphics[width=0.7\textwidth]{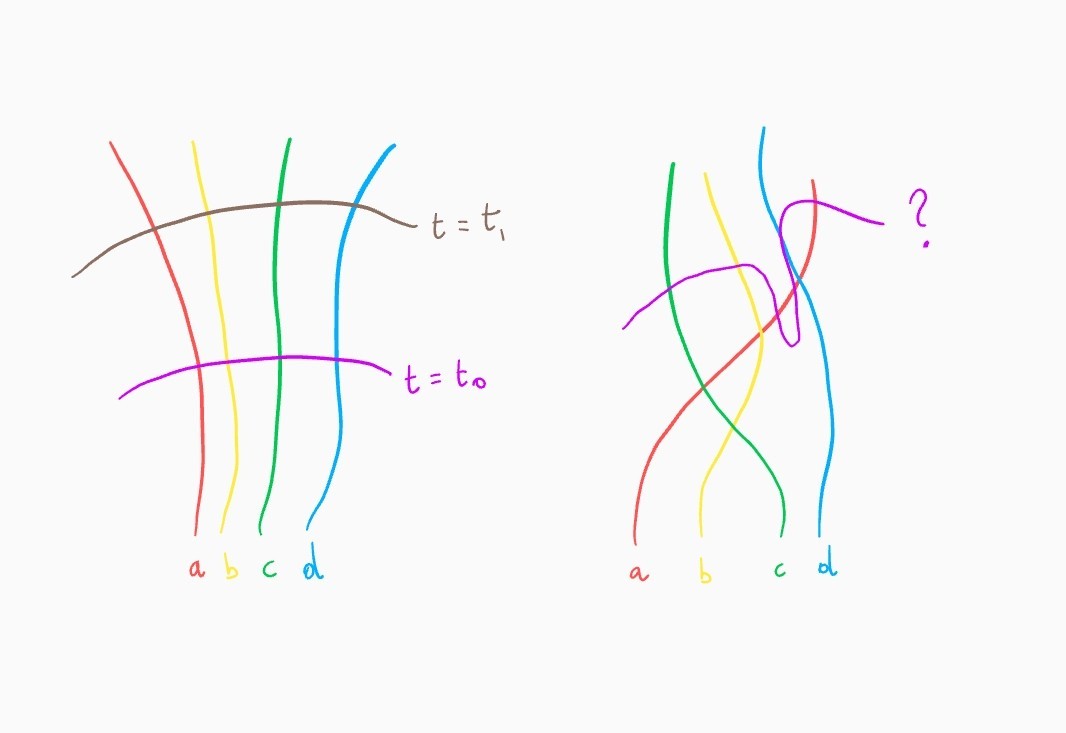}
  \caption*{If the world lines of particles (a,b,c,d) do not intersect and do not twist around each other (zero vorticity) then it is possible to find a family of hypersurfaces orthogonal to them (left). This is impossible if the world lines intersect or fail to form a bundle (right), as in this case there is no global family of hypersurfaces that remains orthogonal to all world lines at all times.}
  \label{fig:weyl}
\end{figure}

\noindent A slightly more technical rephrasing of the statement above is that the world lines of fluid particles should form a bundle of non-intersecting timelike geodesics (a congruence) everywhere orthogonal to a series of spacelike hypersurfaces.

Weyl's postulate translates to a set of geometrical constraints on the metric (\ref{metric}). The particle world lines are defined by $\vec{x} = \text{constant}$, while the spacelike hypersurfaces are defined by $t = \text{constant}$. Since the two are required to be orthogonal, $g_{0 i} = 0$ necessarily. This globally resolves spacetime into a 3+1 split of space and time.

Furthermore, since the fluid particle world lines are geodesics, 
\begin{equation}
    \frac{d^2 x^i}{ds^2} + \Gamma^i_{\mu \nu} \frac{dx^\mu}{ds} \frac{dx^\nu}{ds} = 0
\end{equation}
is satisfied for $x^i = \text{constant}$, meaning that $\Gamma^i_{00} = 0$. From the definition of the Christoffel symbols and the fact that $g_{0i} = 0$, it immediately follows that $\partial_i g_{00} = 0$, so that $g_{00}$ can only depend on $t$. If we then choose the time coordinate to correspond to proper time along the particle world lines, $dt = ds$, we get $g_{00} = 1$, and we can write the metric as
\begin{equation}\label{cosmicmetric}
    ds^2 = -dt^2 + g_{ij}(t,\vec{x}) dx^i dx^j.
\end{equation}
A universe described by (\ref{cosmicmetric}) admits an unambiguous notion of cosmic time and is therefore called \textit{synchronous}\footnote{Observers comoving with the fluid are also called fundamental observers. Any two of those fundamental observers who start with their clocks synchronized at one slice will have identical proper-time readings at every later (or earlier) slice, hence their clocks remain synchronized at all times.}. In a way, metrics that obey Weyl's postulate allow a partial recovery of the Newtonian concept of absolute time \cite{Smeenk}, which is the only way one can meaningfully talk about the passage of time in a cosmological setting. 

Note that the spatial metric in (\ref{cosmicmetric}) is not necessarily homogeneous and isotropic, and, in fact, there are spacetimes that obey Weyl's postulate but violate the cosmological principle. The latter is retrieved only by imposing as an additional condition that there is no preferred spacelike vector orthogonal to the congruence of timelike curves (namely, the observers). As such, Weyl's postulate is a precondition for the cosmological principle, as the latter requires the existence of a fundamental class of observers who all agree on a universal time.

Note also that Weyl's postulate is composed of two fundamental statements. The first is that the motion of matter is well described by a geodesic congruence of world lines\footnote{A geodesic congruence can be interpreted physically as a family of free-falling particles.}. An immediate consequence of this is that the world lines are non-crossing\footnote{This can of course only be true on average and at large enough scales.}. Given that the geodesics are defined by $x^i = \text{constant}$, intersecting world lines would entail a breakdown of the coordinate system on $\mathcal{M}$, as one would have two different values of the coordinates $x^i$ specifying the same spacetime point.

The second is that the congruence of world lines is hypersurface orthogonal. This effectively selects a preferred foliation of spacetime (the one given by taking the unique hyperslices orthogonal to the congruence) and, consequently, a preferred time direction, which is identified with cosmic time.

There is an alternative geometrical characterization of Weyl's postulate that can be used to derive the metric (\ref{cosmicmetric}). Let $u^\mu = dx^\mu/ds$ be the four-velocity field of the matter particles, with $u^\mu u_\mu = -1$. The acceleration vector can be written as
\begin{equation}
    a^\mu = u^\nu \nabla_\nu u^\mu,
\end{equation}
or
\begin{equation}
    (a^\mu u_\nu+\nabla_\nu u^\mu)u^\nu = u^\nu \nabla_\nu u^\mu - a^\mu = 0,
\end{equation}
where the first equality follows from $u^\mu u_\mu = -1$. The identity above shows that the term in parenthesis on the left is orthogonal to the four-velocity field, meaning that it is the transverse part of the covariant derivative $\nabla_\nu u^\mu$. 

We define the projection tensor into the instantaneous rest space orthogonal to $u^\mu$ as
\begin{equation}
    h_{\mu \nu} = g_{\mu \nu} + u_\mu u_\nu.
\end{equation}
This can also be seen as the metric tensor of the hypersurface orthogonal to the four-velocity field. Therefore, by definition, 
\begin{equation}\label{projectedcov}
    a_\mu u_\nu+\nabla_\nu u_\mu = h^\alpha_\mu h^\beta_\nu \nabla_\beta u_\alpha.
\end{equation}
Next, we can decompose (\ref{projectedcov}) into its symmetric and antisymmetric parts:
\begin{equation}\label{symant}
    a_\mu u_\nu+\nabla_\nu u_\mu \equiv \theta_{\mu \nu} + \omega_{\mu \nu},
\end{equation}
where\footnote{As usual, round brackets around indices denote symmetrization, while square brackets denote antisymmetrization, so that $\nabla_{(\beta} u_{\alpha)} = 1/2( \nabla_{\beta} u_{\alpha} + \nabla_{\alpha} u_{\beta})$, and $\nabla_{[\beta} u_{\alpha]} = 1/2( \nabla_{\beta} u_{\alpha} - \nabla_{\alpha} u_{\beta})$.}
\begin{equation}
\begin{split}
    \theta_{\mu \nu} &= h^\alpha_\mu h^\beta_\nu \nabla_{(\beta} u_{\alpha)}, \\
    \omega_{\mu \nu} &= h^\alpha_\mu h^\beta_\nu \nabla_{[\beta} u_{\alpha]}
\end{split}
\end{equation}
are known as the expansion tensor and vorticity tensor, respectively. Because these tensors live on the spatial hypersurface orthogonal to $u^\mu$, they can be thought of as three-dimensional second rank tensors. Therefore, we can decompose the expansion tensor $\theta_{\mu \nu}$ into a trace component plus a traceless component:
\begin{equation}\label{sigmatheta}
    \theta_{\mu \nu} = \sigma_{\mu \nu} + \frac{1}{3} \theta h_{\mu \nu},
\end{equation}
where $\sigma_{\mu \nu}$ is the shear tensor, $\theta \equiv h^{\mu \nu} \theta_{\mu \nu}$ is the trace of the expansion tensor (expansion scalar), and the factor of 1/3 is related to the dimensionality of the projected space. The vorticity tensor is antisymmetric, so automatically traceless. 

Using (\ref{symant}) and (\ref{sigmatheta}), we can write the kinematical decomposition of the timelike congruence as
\begin{equation}\label{kin}
    \nabla_\nu u_\mu = \sigma_{\mu \nu} + \omega_{\mu \nu} + \frac{1}{3} \theta h_{\mu \nu} - a_\mu u_\nu.
\end{equation}
Physically, the expansion scalar measures how the volume of a small cloud of particles changes over time, the shear tensor describes how the cloud changes shape (distorts) without changing its volume, while the vorticity tensor captures the tendency of nearby particles to rotate or swirl around each other.

Crucially, the vorticity tensor vanishes if and only if the world lines in the congruence are everywhere orthogonal to the spatial hypersurface in some foliation of spacetime, i.e. if and only if Weyl's postulate holds \cite{Malament}. To show this, first note that the field $u^\mu$ is hypersurface orthogonal if it can be written as $u_\mu = \nabla_\mu f$ for some smooth real-valued function $f$. This is because the hypersurfaces of constant $f$ are everywhere orthogonal to the gradient.

If $u_\mu = \nabla_\mu f$ then 
\begin{equation}
    \omega_{\mu \nu} = h_\mu^\alpha h_\nu^\beta \nabla_{[\beta} \nabla_{\alpha]} f = 0
    \label{w=0}
\end{equation}
since the metric connection is assumed to be torsion-free\footnote{If the connection is torsion-free, the Christoffel symbols are symmetric, $\Gamma_{\mu \nu}^\alpha = \Gamma_{\nu \mu}^\alpha$, meaning that $\nabla_\mu \nabla_\nu f = \nabla_\nu \nabla_\mu f = \partial_\mu \partial_\nu f - \Gamma_{\mu \nu}^\alpha \nabla_\alpha f$.}. The converse implication, namely that if the rotation field vanishes everywhere then the four-velocity is locally hypersurface orthogonal, is not as immediate, and is in fact a special case of Frobenius' theorem \cite{Wald}. Having established the equivalence of Weyl's postulate with vanishing vorticity everywhere, the metric (\ref{cosmicmetric}) follows naturally. 

Geometrically, the conditions $\omega_{\mu \nu} = 0$, $u_\mu = \nabla_\mu f$, and $\nabla_{[\mu} u_{\nu]} = 0$ are all equivalent, and characterize a geodesic congruence that is hypersurface orthogonal. Intuitively, a non-zero vorticity would make the family of timelike curves (fibers) twist around in such a way that, by pointing in different directions, it would be impossible to slice it so that each slice is orthogonal to all fibers. 

Having determined the geometric conditions that make it possible for time to have a direction in the cosmological setting, let us now try to deduce further geometric constraints on the metric (\ref{cosmicmetric}) such that time can also have an orientation.

\section{The Past Hypothesis}\label{past}
How do we give time an orientation, such that we can distinguish between forward and backward time evolution? There is nothing in Einstein's equations that can be used to pick up a preferred time orientation, thus we must look at some macroscopic feature of the world that lies outside the microphysics of gravitation. 

This feature has its roots in thermodynamics \cite{Albert}; specifically, the distribution of matter and energy in the universe, which has a very natural characterization in terms of spacetime geometry, as we will see\footnote{Naturally, the distribution of matter and energy in the universe fully determines the spacetime geometry, so the thermodynamic picture and the geometrical picture are equivalent in GR.}. What we need is a measurable macroscopic quantity that changes \textit{monotonically} with the cosmic time $t$. If such physical quantity exists, we can then exploit its asymmetry to define a direction of time on $\mathcal{M}$. This quantity is usually identified with the entropy $S$ of the matter fields that live on $\mathcal{M}$, plus that of gravity. By convention, one identifies the direction of increasing time with that of increasing entropy.

Entropy can only monotonically increase in one direction if it was initially low. This is the content of the Past Hypothesis \cite{Albert2}:
\begin{quote}
\textbf{Past Hypothesis}\\
The universe began in a special low-entropy state.
\end{quote}
The formulation of the past hypothesis requires the existence of a ``beginning'', i.e. a particular boundary in spacetime on which we set our initial condition of low entropy. Such a boundary is naturally given by the initial cosmological singularity, which is a generic prediction of General Relativity given the strong energy condition holds \cite{Hawking2}. 

\begin{figure}
  \centering
  \includegraphics[width=0.7\textwidth]{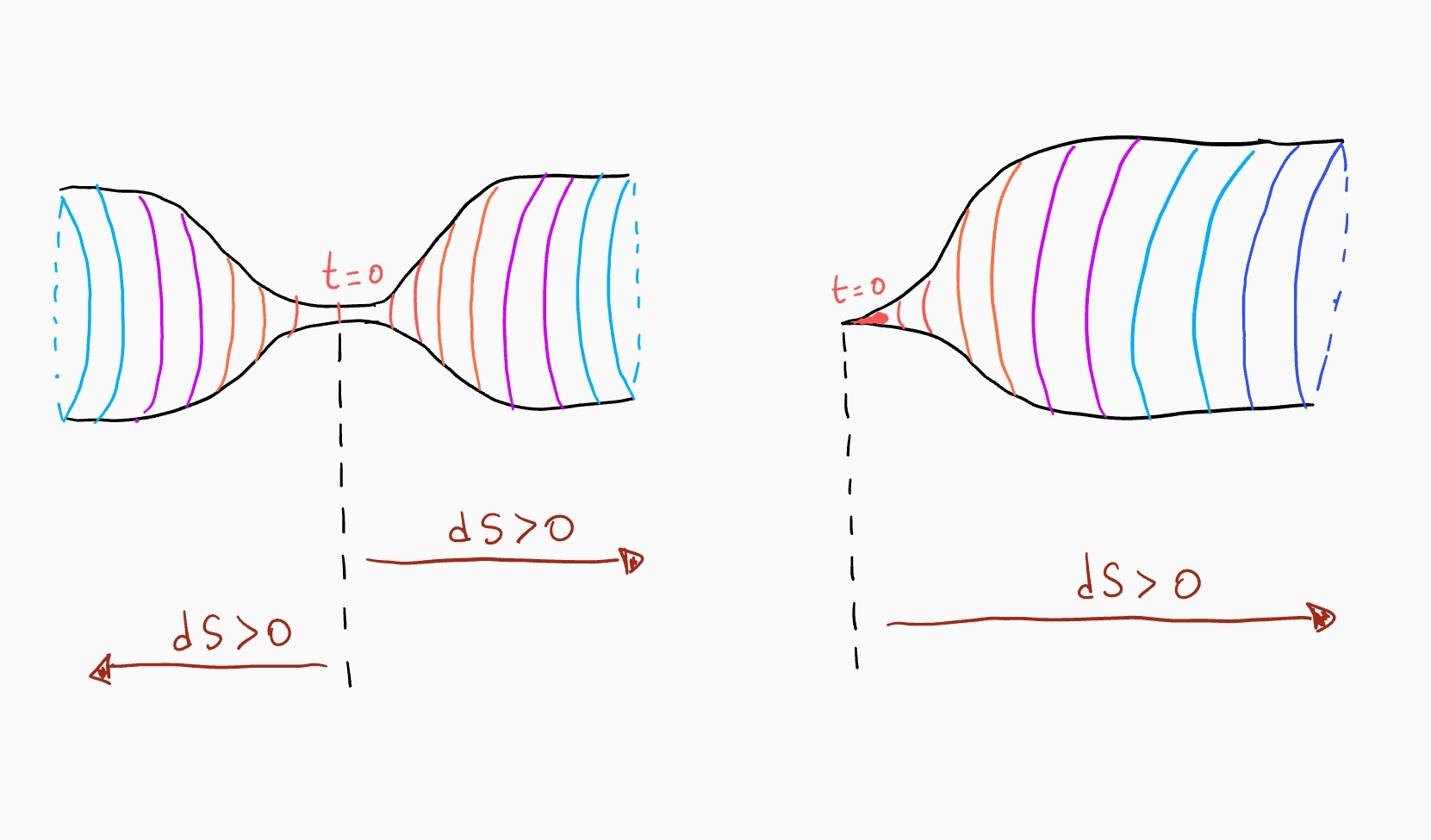}
  \caption*{In a Big Bounce scenario (left) one can impose the regularity condition (\ref{PHW}) to ensure that the gravitational entropy is negligible at $t=0$. However, given the spacetime is past-extendible, entropy would then be expected to increase in both directions of time, and neither could be chosen as proceeding from past to future. On the other hand, if spacetime is non-extendible (right), as in the case of a cosmological singularity, there is only one branch of thermodynamic evolution possible, which implicitly determines the ``arrow of time'', i.e. time's orientation from past (Big Bang) to future.}
  \label{fig:past}
\end{figure}

To see this, recall equation (\ref{kin}). By the definition of the Riemann tensor, we can write
\begin{equation}
    (\nabla_\beta \nabla_\nu u_\mu - \nabla_\nu \nabla_\beta u_\mu) u^\beta = R_{\mu \sigma \nu \delta} u^\sigma u^\delta.
\end{equation}
Using (\ref{kin}), and after a bit of algebra, we get
\begin{equation}\label{Beldecom}
    \frac{2}{3} \theta \sigma_{\mu \nu} - \sigma_{\mu \alpha} \sigma^\alpha_\nu - \omega_{\mu \alpha} \omega^\alpha_\nu - \frac{1}{3} \left( \dot{\theta} + \frac{\theta^2}{3}\right) h_{\mu \nu} - h_\mu^\alpha h_\nu^\beta \left(\dot{\sigma}_{\alpha \beta} - \nabla_{(\beta} a_{\alpha)}\right) - a_\mu a_\nu = R_{\mu \sigma \nu \delta} u^\sigma u^\delta,
\end{equation}
where the dot indicates differentiation with respect to cosmic time. Taking the trace of (\ref{Beldecom}) for $a_\mu = 0$ (geodesic motion) gives Raychaudhuri's equation \cite{Tahim}:
\begin{equation}
    \dot{\theta} = \omega^2 - \sigma^2 - \frac{\theta^2}{3} - R_{\mu \sigma \delta}^\mu u^\sigma u^\delta,
\end{equation}
with $\omega^2 = \omega_{\mu \nu} \omega^{\mu \nu}$ and $\sigma^2 = \sigma_{\mu \nu} \sigma^{\mu \nu}$. When the vorticity is zero (\S \ref{Weyl}), Raychaudhuri's equation becomes
\begin{equation}\label{Ray}
    \dot{\theta} = - \sigma^2 - \frac{\theta^2}{3} - R_{\mu \nu} u^\mu u^\nu,
\end{equation}
where $R_{\mu \nu} \equiv R^\alpha_{\mu \alpha \nu}$ is the Ricci tensor. 

Thus, if one assumes the strong energy condition (SEC) $R_{\mu \nu} u^\mu u^\nu \geq 0$, the right hand side of (\ref{Ray}) is always negative, leading to the inequality
\begin{equation}
    \dot{\theta} \leq - \frac{\theta^2}{3}.
\end{equation}
Integrating the inequality with respect to cosmic time gives
\begin{equation}
    \frac{1}{\theta} \geq \frac{1}{\theta_0} + \frac{t}{3}.
\end{equation}
If $\theta_0 <0$, then the expansion scalar diverges in a (finite) cosmic time of at most $t=3/|\theta_0|$. This is known as the focusing theorem. The focusing theorem is purely local: if the expansion scalar $\theta$ of a timelike geodesic congruence ever becomes negative, then $\theta \rightarrow -\infty$ within finite proper time. If in addition there exists a spacelike Cauchy hypersurface $\Sigma$ on which $\theta|_\Sigma < 0$ everywhere, namely if there exists a uniformly contracting Cauchy slice, then one can show that every geodesic has to terminate after finite proper time \cite{HawkingPenrose}, i.e. the spacetime is (past) timelike geodesically incomplete, which is the definition of a cosmological singularity.

The situation is slightly different for accelerating spacetimes with $\Lambda >0$ \cite{Borde, BGV, Galloway}. In this case the SEC needs to be replaced by the weaker condition
\begin{equation}
    R_{\mu \nu} u^\mu u^\nu \geq 3 H^2 g_{\mu \nu} u^\mu u^\nu,
    \label{weakSEC}
\end{equation}
where $H^2=\Lambda/3$. As we saw, the metric admits a preferred slicing with timelike four-velocity field $u^\mu = (1,0,0,0)$, therefore (\ref{weakSEC}) is simply
\begin{equation}
    R_{00} \geq - 3H^2 .
    \label{R_00}
\end{equation}
Then, (\ref{Ray}) gives the bound
\begin{equation}
    \dot{\theta} \leq -\frac{\theta^2}{3} + 3 H^2.
\end{equation}
The right-hand side is now negative only if $|\theta|$ is large enough. In fact, focusing happens only when 
\begin{equation}
    \theta|_\Sigma < - 3 H,
    \label{theta}
\end{equation}
everywhere on a compact Cauchy hypersurface $\Sigma$. 

What further constraint does this place on the metric (\ref{cosmicmetric})? The conditions (\ref{R_00}) and (\ref{theta}) effectively select a subset of all possible cosmic-time metrics (\ref{cosmicmetric}). To see this, let us consider the simple case of a homogeneous and isotropic universe. The line-element is given then by the FRW metric:
\begin{equation}\label{FRW}
    ds^2 = -dt^2 + a^2(t) \left[ \frac{dr^2}{1-kr^2} + r^2 d\Omega^2\right], 
\end{equation}
where $a(t)>0$ is the scale factor and $k=-1, 0 , 1$ the curvature parameter. Now, one can show that for the metric (\ref{FRW}), $R_{00} = -3 \ddot{a}/a$, therefore $R_{00} \geq -3H^2$ whenever $\ddot{a} \leq H^2 a$ for all values of $t$. Effectively, this selects universes in which the acceleration is never larger than the de Sitter value.

Exotic matter or quantum effects that violate the assumptions above can give rise to cosmological models that have no beginning. For example, the simple bouncing model described by 
\begin{equation}\label{bouncing}
    a(t) = a_b \left[ 1 + \left( \frac{t}{t_b}\right)^2\right]^q, \quad q>0,
\end{equation}
can be shown to arise due to quantum cosmological effects described by the Wheeler-DeWitt equation \cite{Pinto} and has no cosmological singularity. In fact, in this model $\ddot{a} \leq 0$ only when $|t/t_b| \geq 1/\sqrt{1-2q}$ with $0<q<1/2$. This means that around the bounce, i.e. for $|t/t_b| \leq 1/\sqrt{1-2q}$, the universe reverses course, starts expanding, and the singularity is avoided. In this region $\ddot{a}/a = 2 q/t_b^2 \gg \Lambda$, so condition (\ref{R_00}) is violated.

Out of the class of spacetimes described by (\ref{cosmicmetric}) we thus only take the ones that are past geodesically incomplete, so that we may impose our initial condition on entropy on a suitable (singular) spacelike boundary. Note that this could in principle be done also on a non-singular (past-extendible) spacelike boundary, like the bounce of metric (\ref{bouncing}). However, the entropy would then be expected to increase in \textit{both directions} of time, with the result that we could not pick a preferred one. It is then crucial that our spacelike boundary be inextendible, so that there is only one thermodynamic branch of evolution available to the universe.

Given that such a singularity exists, what is the geometric condition that has to be satisfied in order for it to represent a low-entropy state? A famous proposal is the Weyl curvature hypothesis \cite{Penrose2} . Penrose's hypothesis is that at the initial cosmological singularity the conformal part of the Riemann curvature should vanish (or be negligible compared with the Ricci part). Typically, this is expressed by demanding that any dimensionless scalar built from the Weyl tensor goes to zero as cosmic time $t \rightarrow 0$. A concrete proposal for the ``gravitational entropy'' is \cite{Wainwright, Goode}
\begin{equation}\label{PHW}
    \lim_{t \rightarrow 0} \frac{C_{\alpha \beta \gamma \delta} C^{\alpha \beta \gamma \delta}}{R_{\mu \nu}R^{\mu \nu}} = 0,
\end{equation}
where $C_{\alpha \beta \gamma \delta}$ is the Weyl tensor and $R_{\mu \nu}$ the Ricci tensor. Condition (\ref{PHW}) may be viewed as the precise, geometric content of the past hypothesis: gravitational entropy starts essentially at zero because the tidal part of the field is initially suppressed. 

This requirement of no ``clumping'' effectively forces the initial singularity to be regular, or isotropic, in the sense of \cite{Goode}. This means that the spatial metric must admit a conformally flat expansion, 
\begin{equation}\label{expansion}
    g_{ij}(t, \vec{x}) = a^2(t) \left[ \gamma_{ij}^{(0)}(\vec{x}) + t^n \gamma_{ij}^{(1)}(\vec{x}) + ...\right], \quad n>0,
\end{equation}
where the leading 3-metric $\gamma_{ij}^{(0)}(\vec{x})$ is a space of constant sectional curvature. This condition ensures the vanishing of the Weyl tensor at the singular boundary, since conformal flatness implies \( C_{\alpha\beta\gamma\delta} = 0 \). Indeed, the full FRW metric (\ref{FRW}) satisfies this identically for all \( k = \pm 1, 0 \), and thus provides a prototypical example of an isotropic singularity consistent with the Weyl curvature hypothesis.

To summarize, if conditions (\ref{w=0}), (\ref{R_00}), and (\ref{theta}) are satisfied, and $g_{ij}(t,\vec{x}) \sim a^2(t) \gamma_{ij}^{(0)}(\vec{x})$ for $t \rightarrow 0$, the universe admits a notion of cosmic time with a well-defined direction (that orthogonal to the hypersurfaces of constant cosmic time) and orientation (proceeding from the low-entropy initial singularity). While this ensures the existence of an arrow of time on a local patch, there could still be topological effects that spoil chronology globally, as we will see in the next two sections.

\section{Time-orientability}\label{Mobius}
The metric (\ref{cosmicmetric}) does not fix the global topology of the spacetime manifold $\mathcal{M}$. Assuming $\mathcal{M}$ is a smooth manifold and $\Gamma$ a discrete group of isometries acting smoothly, freely, and properly on $\mathcal{M}$, the quotient space $\tilde{\mathcal{M}}=\mathcal{M}/\Gamma$ is again a smooth manifold with the same local geometry as $\mathcal{M}$ \cite{Lee}. Clearly, $\tilde{\mathcal{M}}$ and $\mathcal{M}$ are locally isometric but may differ globally in topological and causal structure. In such cases, $\mathcal{M}$ is called the covering space of $\tilde{\mathcal{M}}$.

Let's take the usual FRW universe as a simple example. The metric (\ref{FRW}) can be written more generally as\footnote{We use here the notation of (\ref{expansion}).}
\begin{equation}\label{FRW+}
    ds^2 = -dt^2 + a^2(t) \gamma_{ij} dx^i dx^j, \quad t>0,
\end{equation}
where the scale factor $a(t)$ is defined only for positive times. We denote the manifold corresponding to the metric (\ref{FRW+}) with $\mathcal{M}_+$. The time-reversed solution $a(-t)$ defined for $t<0$ describes an isometric copy of $\mathcal{M}_+$, which we call $\mathcal{M}_-$.

Now we can choose as our covering space $\mathcal{M} = \mathcal{M}_+ \cup \mathcal{M}_-$\footnote{A disjoint union of two manifolds is a manifold.}, equipped with the FRW metric on each side, and with topology $\mathcal{M} = (\mathbb{R} \setminus \{0\}) \times \Sigma$, where the singularity is excised from the manifold and $\Sigma$ is the FRW 3d spatial slice. We then introduce the fixed-point-free isometry $\gamma:\mathcal{M} \rightarrow \mathcal{M}$, defined by
\begin{equation}
    \gamma(t,\vec{x}) = (-t, \Omega \vec{x}),
\end{equation}
where $\vec{x} \rightarrow \Omega \vec{x}$ is an isometry of the spatial slice $\Sigma$. For example, if $\Sigma = \mathbb{R}^3$, $\Omega$ can be a translation $\vec{x} \rightarrow \vec{x} + \vec{a}$ by a constant vector $\vec{a}$. The cyclic group generated by this isometry is $\Gamma = \langle \gamma^n \, | \, n\in \mathbb{Z} \rangle \cong \mathbb{Z}$. That is, we identify
\begin{equation}\label{mobiusid}
    (t,\vec{x}) \sim \left( (-1)^n t, \vec{x} + n \vec{a} \right) \quad \forall n \in \mathbb{Z}.
\end{equation}
Since $\gamma$ is smooth and fixed-point-free, and the action of $\Gamma$ is free and properly discontinuous, the quotient space $\tilde{\mathcal{M}} = \mathcal{M}/\Gamma$ is a smooth manifold that is locally isometric to FRW but globally non-time-orientable. The resulting spacetime resembles a twisted cylinder in the time direction: following a closed timelike path through successive identifications causes the time orientation to reverse.

\begin{figure}
  \centering
  \includegraphics[width=0.4\textwidth]{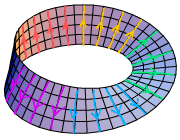}
  \caption*{The universe could have sections that look like a Möbius strip, as in the picture, with time going vertically and space horizontally. Going around the strip once would flip the direction of time, with consequent global loss of orientability.}
  \label{fig:mobebius}
\end{figure}

At this point, it is useful to remind oneself of the definition of a time-orientable manifold \cite{HawkingEllis}:
\begin{quote}
\textbf{Time-orientability}\\
A spacetime $\mathcal{M}$ is said to be time-orientable if there exists a continuous, globally defined timelike vector field $v^\mu$ such that at every point $p \in \mathcal{M}$, $v^\mu(p)$ is future-directed.
\end{quote}
Equivalently, a Lorentzian manifold is time-orientable if one can make a globally consistent choice of ``future'' and ``past'' cones at every point in the manifold. If such a choice cannot be made globally, i.e. if transporting a timelike vector around a closed loop causes it to flip its time direction, then the manifold is not time-orientable \cite{Geroch}.

Note then that under the action of $\gamma$, a future-directed timelike vector at $(t,\vec{x})$ is mapped to a past-directed timelike vector at the point $\gamma(t,\vec{x})$. Since these two points are identified in $\tilde{\mathcal{M}}$, there is no consistent way to define ``future'' and ``past'' globally. In other words, parallel transporting a future-directed timelike vector around a closed loop in $\tilde{\mathcal{M}}$ that lifts to a path in $\mathcal{M}$ starting at $(t,\vec{x})$ and ending at $\gamma(t,\vec{x}) = (-t, \vec{x}+\vec{a})$ flips its time direction.

In such a universe one might be able to distinguish past from future locally, but not globally, so while time might be a well-defined concept for local observers, it loses its meaning entirely for sufficiently adventurous space travelers. 

\section{Chronology}\label{chronology}
If the universe is both synchronous and time-orientable, we have a global notion of time orientation, i.e. we can distinguish \textit{globally} between  ``past'' and ``future''. In order to be able to extend the local arrow of time to its global instantiation, however, we need to make sure that the universe is also chronological, namely that there is a well-defined \textit{ordering} of events. 

If one defines the chronological relation as $x$ chronologically precedes $y$, which we denote by $x \prec y$, iff there exists a future-directed timelike curve from $x$ to $y$, then requiring chronology means requiring that the chronological relation is irreflexive, meaning that $p \nprec p$ $\forall p \in \mathcal{M}$. Note that were this false, and all other conditions listed so far true (synchronicity, time-orientability, initial regularity, etc.), we would still fail to have a global arrow of time, as entropy could not increase monotonically with cosmic time forever and for all observers.

The absence of a chronological order of events is the result of closed timelike curves \cite{Luminet}. A closed timelike curve (CTC) is a continuous timelike curve whose initial and final points coincide. Therefore,
\begin{quote}
\textbf{Chronology}\\
A spacetime $\mathcal{M}$ is called chronological if it contains no closed timelike curves.
\end{quote}
We already assumed the spacetime $\mathcal{M}$ to be synchronous, and as such described by the metric (\ref{cosmicmetric}). Effectively, this means the universe admits a time function \cite{Beem}, namely a smooth function $t:\mathcal{M} \rightarrow \mathbb{R}$ such that $\nabla t$ is everywhere timelike, and the level sets $\Sigma = \{ p \in \mathcal{M} | t(p) = \text{const}\}$ are spacelike hypersurfaces. The function $t$ increases strictly on every future-directed causal (timelike and null) curve and is known as cosmic time. This is a mere paraphrasis of what was said in \S \ref{Weyl}: a synchronous spacetime can be foliated into a family of spacelike hypersurfaces indexed by a universal time coordinate. 

Suppose now $\gamma$ is a future-directed timelike curve from a point $p$ back to itself (a CTC). Then the value of $t$ must strictly increase along $\gamma$, implying $t(p) > t(p)$, a contradiction. Hence, no closed timelike curves can exist in a synchronous spacetime. What's more, Hawking proved that the existence of any time function is equivalent to the spacetime being \textit{stably causal}, i.e. CTCs cannot even be created by any small perturbation of the metric \cite{Hawkingfunc}.

Stable causality is not yet the top of the causal ladder \cite{Minguzzi}. The strongest condition is global hyperbolicity: roughly speaking, a spacetime manifold is called globally hyperbolic if the future state of the system is specified by initial conditions \cite{Beem}. Geroch showed that global hyperbolicity is equivalent to the existence of a Cauchy surface on $\mathcal{M}$ \cite{Geroch2}, meaning that the manifold is topologically equivalent to $\mathbb{R} \times \Sigma$ for some Cauchy surface $\Sigma$. We saw in \S \ref{Weyl} that synchronicity requires the congruence to be geodesic and irrotational, and consequently that all causal curves intersect each constant time hypersurface $\Sigma$ exactly once. This ensures that $\Sigma$ is a Cauchy surface, and the spacetime manifold is globally hyperbolic, the strongest of the causal conditions, which in turn guarantees a well-posed initial-value formulation of Einstein's equations.

Weyl's postulate ensures the spacetime manifold is causally well-behaved, but could there be non-trivial topological identifications, like the ones in \S \ref{Mobius}, that spoil chronology? Not really, as we will now see. 

In \S \ref{Mobius} we effectively created a Möbius strip in spacetime by identifying, for example, the spacetime points $(t, \vec{x}) \sim (-t, \vec{x} + \vec{a})$. This is clearly an isometry of (\ref{FRW+}), so the construction $\mathcal{M}/\Gamma$ is possible\footnote{Note that the transformation (\ref{mobiusid}) is not necessarily an isometry of (\ref{cosmicmetric}). In \S \ref{Mobius} we were merely interested in providing a concrete counter-example to time-orientability in the class of synchronous spacetimes.}. An analogous topological identification meant to create CTCs would be of the type
\begin{equation}\label{ctcid}
    (t, \vec{x}) \sim (t+a, \vec{x}).
\end{equation}
However, for (\ref{ctcid}) to define a proper isometry, the metric must be invariant under time translations, which (\ref{cosmicmetric}) cannot be in any case, given the requirement in (\ref{expansion}). Roughly speaking, a universe with a well-defined notion of (thermodynamic) time must be evolving in (coordinate) time and as such cannot be invariant under $t \rightarrow t + a$. 

\begin{figure}
  \centering
  \includegraphics[width=0.9\textwidth]{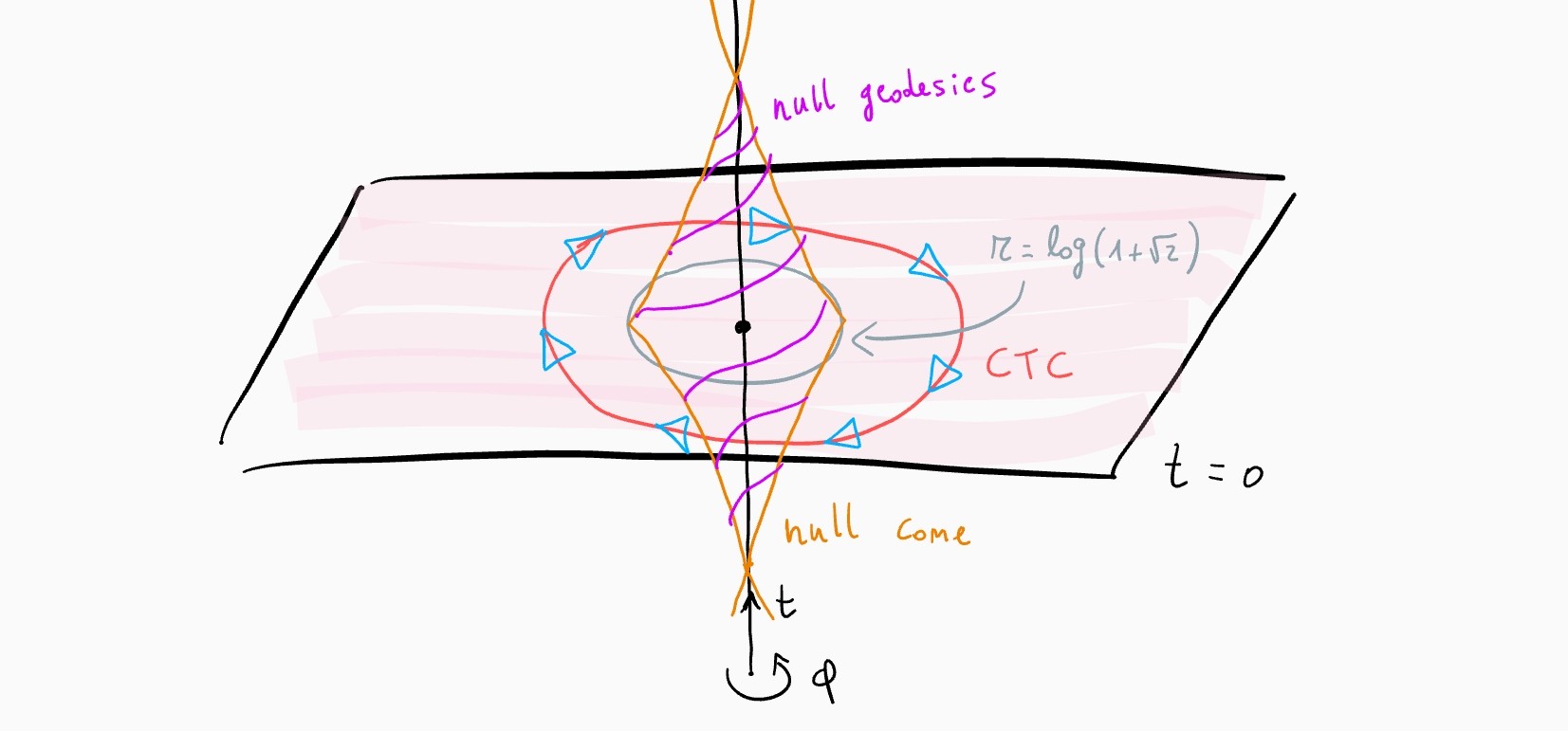}
  \caption*{Spacetimes with non-zero vorticity, such as Gödel's universe, may contain closed timelike curves (CTCs, shown in red). In these spacetimes, the light cones (in blue) tilt sufficiently to allow world lines to loop back on themselves, making time travel to the past theoretically possible. In Gödel's universe, this behavior arises due to the rigid rotation of spacetime around the central axis depicted. Beyond the radius \( r = \log(1 + \sqrt{2}) \), there exists a timelike congruence consisting of circular trajectories.}
  \label{fig:CTC}
\end{figure}

The class of universes we are interested in admits no timelike Killing vector field, hence any identification along a timelike vector field would fail to preserve the metric (it would not be an isometry), and the quotient space $\mathcal{M}/\Gamma$ would not inherit the local geometry of $\mathcal{M}$. Only in exceptional cases, such as the Einstein static universe where $a(t) = \text{const}$, does time translation generate a global isometry. These universes, however, would be devoid of initial singularity, and there would be no way to impose the condition of regularity required by the Past Hypothesis.

Note finally that the requirement of zero vorticity $\omega_{\mu \nu} = 0$ is crucial in ensuring chronology on any local patch of the manifold. A famous counterexample is given by Gödel's universe \cite{Godel}, which is neither synchronous nor chronological, and the reason for both is to ascribe to a non-zero vorticity. Gödel's metric can be written in cylindrical coordinates as
\begin{equation}\label{godel}
    ds^2 = \frac{1}{2\omega^2}\left( -dt^2 + dx^2 - \frac{1}{2} e^{2x} dy^2 + dz^2 - 2e^x dt dy\right),
\end{equation}
where $\omega$ is the angular velocity of the dust about the $y$ axis. The presence of the cross-term $dtdy$ prevents the constant-time hypersurfaces from being orthogonal to the dust world lines, and Weyl's postulate is violated. As we saw in \S \ref{Weyl}, this is equivalent to saying that the vorticity is non-zero, and in fact $\omega_{\mu \nu} = e^x/(2 \sqrt{2} \omega) (\delta^y_\mu \delta_\nu^x-\delta_\mu^x \delta_\nu^y)$, with $\omega^2 \equiv 1/2 \, \omega_{\mu \nu} \omega^{\mu \nu}$ \cite{Barrow}. This also leads to the existence of CTCs: the same frame-dragging term $-2 e^x dt dy$ that spoils synchrony also tips the light cones far enough to let a particle loop back to its own past. To see this, write the metric (\ref{godel}) in polar coordinates $(t,r,\phi,z)$, defined as
\begin{equation}
\begin{aligned}
  e^{x} &= \cosh(2r) + \cos\phi\,\sinh(2r), \\
  y     &= e^{-x}\,\sin\phi\,\sinh(2r), \\
  t     &= 2\sqrt{2}\,\arctan\!\bigl(e^{-2r}\tan\tfrac{\phi}{2}\bigr)
          - \sqrt{2}\,\phi + 2\tau,
\end{aligned}
\end{equation}
where $r \geq 0$ and $0 \leq \phi \leq 2 \pi$. In these variables the metric takes the form \cite{Marquet}
\begin{equation}
    ds^2 = \frac{1}{2\omega^2} \left[ -d\tau^2 + dr^2 +dz^2+(\sinh^2r-\sinh^4r)d\phi^2 - 2 \sqrt{2} \sinh^2 r d\tau d\phi\right].
\end{equation}
Consider now the obvious ring curve 
\begin{equation}
    \gamma : \quad r=r_0=\text{const}, \quad z=\text{const}, \quad \tau = \tau_0 = \text{const}, \quad \phi \in [0, 2\pi].
\end{equation}
Its tangent vector is simply $v^\mu = \partial_\phi^\mu$, so the squared norm along $\gamma$ is the $\phi \phi$ component of the metric:
\begin{equation}
    v^2 = g_{\phi \phi}(r_0) = \frac{1}{2 \omega^2} \sinh^2 r_0 (1-\sinh^2r_0).
\end{equation}
Given $f(r) = \sinh^2r(1-\sinh^2r)$, $f(r) = 0$ when $\sinh r = 1$, i.e. for $r=r_c=\sinh^{-1}1 \approx 0.88$. This means that for $r>r_c$, one has $f(r)<0$, thus $v^2<0$, meaning that the ring is timelike. Because $\phi$ is intrinsically periodic, the timelike direction closes on itself: every point with $r>r_c$ lies on a closed timelike curve of length $2 \pi$\footnote{Note that Gödel's universe's violation of chronology is a geometrical, rather than topological, effect. In fact, the 4-manifold underlying Gödel's solution is just $\mathbb{R}^4$; it is simply-connected and has no exotic identifications. Its strangeness (frame-dragging, tipped light-cones, closed timelike curves) comes from the metric coefficients, not the manifold's topology.}.

The requirement that $\omega_{\mu \nu} = 0$ everywhere on the manifold is enough to guarantee the absence of CTCs on any local patch. Furthermore, the necessity of having a universe that expands from an initial singularity in order to satisfy the past hypothesis (\S \ref{past}) prevents any topological identification of the type (\ref{ctcid}) that would create CTCs globally. We conclude then that our class of universes automatically satisfies chronology in all realistic scenarios.

\section{The Role of Entropy}\label{entropy}
Castagnino and Lombardi, following Earman's heresy, set out to define a global arrow of time for the universe independent of entropy \cite{Castagnino2}. Their argument can be summarized as follows. It is usually assumed that time asymmetry can only be retrieved from a time-non-invariant law. Such is the second law of thermodynamics, and that is the chief reason why this law has been chosen as determining the arrow of time, since all other microscopic physical laws are indeed time-invariant.

However, even when the fundamental dynamical laws are time-invariant, their solutions need not be. This is an instance of symmetry breaking \cite{Prigogine}: while each individual solution may be asymmetric, the full symmetry of the equations of motion is restored when one considers the set of all possible solutions. Whenever there is a physical process that realizes only one particular element of the set, the symmetry will be broken.

According to this view, the universe can be asymmetric in time even though the fundamental laws of physics are time-reversal invariant, and so the arrow of time can be defined in purely geometrical terms without the need of external non-temporal variables. As a simple example, take the Friedmann equation for a FRW universe:
\begin{equation}\label{fried}
    \left(\frac{\dot{a}}{a} \right)^2 = \frac{8 \pi G}{3} \rho - \frac{k}{a^2} + \frac{\Lambda}{3}.
\end{equation}
This equation is time-reversal invariant because acting with the transformation $t \rightarrow -t$ leaves it unchanged. Solving for a flat ($k = 0$), matter-dominated universe we get two solutions:
\begin{equation}\label{expcon}
\begin{split}
    a_{\text{exp}}(t) = a_0 \left( \frac{t}{t_0} \right)^{2/3} &\text{for } t>0, \\
    a_{\text{con}}(t) = a_0 \left( -\frac{t}{t_0} \right)^{2/3} &\text{for } t<0.
\end{split}
\end{equation}
The first describes an expanding universe, while the second describes a contracting universe. Both are perfectly legitimate solutions of the dynamical equations, but only one of them can be physically realized\footnote{Note again that it is the singularity at $t=0$ that prevents the two solutions from being glued together to restore symmetry, reinforcing the idea that a singularity is needed to define an arrow of time, see \S \ref{past}.}. Whichever one chooses, time-reversal invariance will be broken.  

Consider instead a matter-dominated spatially closed ($k=+1$) universe. The solution of (\ref{fried}) can be given parametrically as
\begin{equation}\label{cycloid}
\begin{split}
    a(\eta) = \frac{A}{2}\left( 1 - \cos \eta \right), \\
    t(\eta) = \frac{A}{2}\left( \eta - \sin \eta \right),
\end{split}
\end{equation}
where $0<\eta<2\pi$ is the conformal time parameter, and $A\equiv 8 \pi G /3 \rho_0 a_0^3$. In (\ref{cycloid}), $\eta = 0$ corresponds to the Big Bang ($a=0$), $\eta = \pi$ corresponds to the period of maximum expansion of the universe ($\dot{a}=0$), and $\eta = 2\pi$ corresponds to the Big Crunch ($a=0$ again). Notice now that, contrary to (\ref{expcon}), (\ref{cycloid}) is symmetric. Under the transformation $\eta \rightarrow 2 \pi - \eta$, $a(2\pi-\eta) = a(\eta)$ and $t(2\pi-\eta)=2t_{\text{max}}-t(\eta)$, where $t_{\text{max}}=A\pi/2$. Setting $\tilde{t}\equiv t-t_{\text{max}}$, this mapping thus sends $\tilde{t} \rightarrow-\tilde{t}$ while leaving $a$ unchanged. Hence $a(\tilde{t})$ is an even function about the turnaround and, at least mathematically, the expansion phase is the time-reverse of the contraction phase.

\begin{figure}
  \centering
  \includegraphics[width=0.9\textwidth]{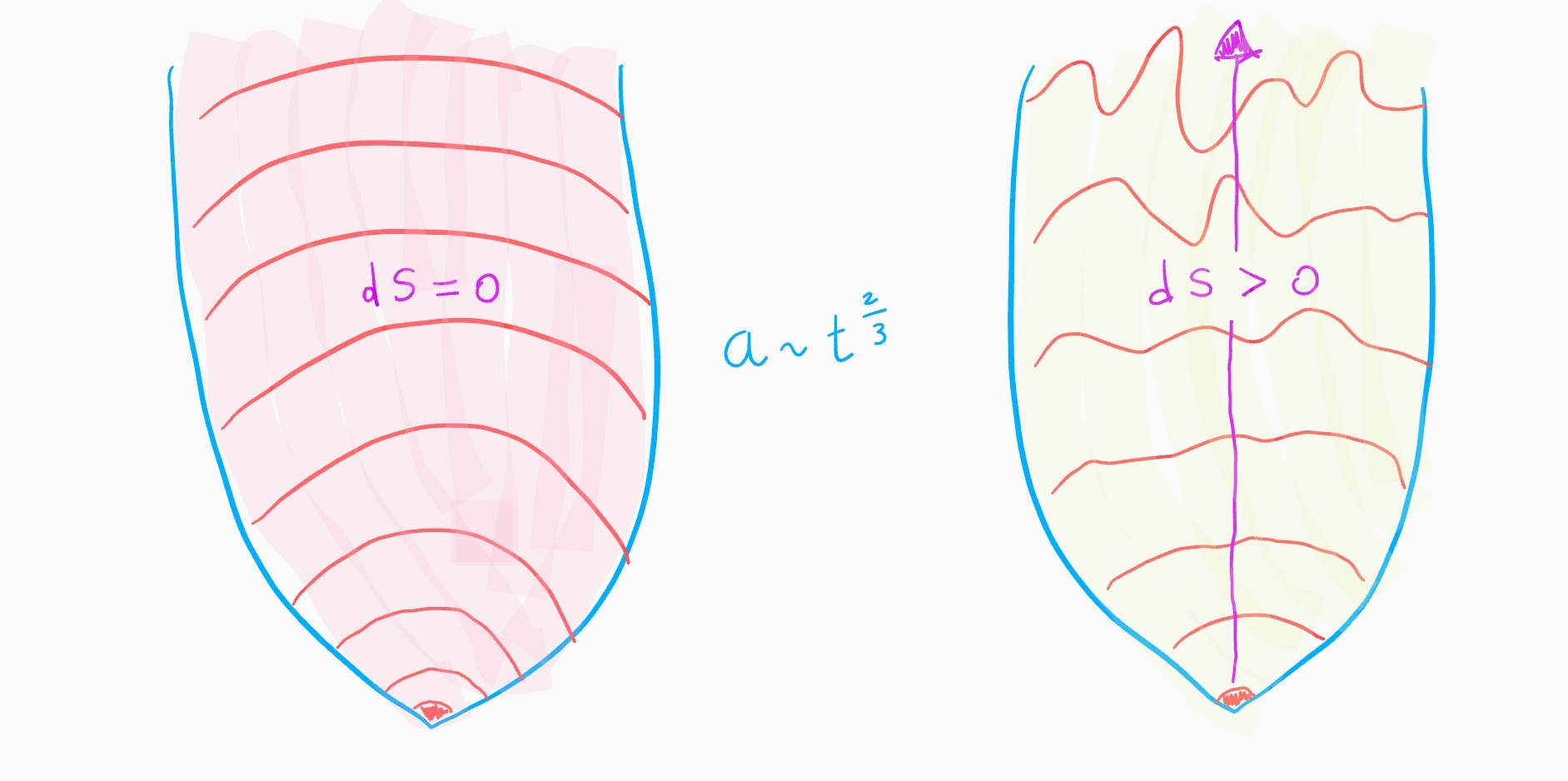}
  \caption*{The spacetime on the left describes a perfectly homogeneous, flat, matter-dominated FRW universe, while the one on the right a flat, matter-dominated FRW universe with small initial inhomogeneities. Both spacetimes are temporally asymmetric but only the one on the right will display an arrow of time; one that proceeds from the regular Big Bang to the irregular future with a more lumpy matter distribution.}
  \label{fig:entropy}
\end{figure}

According to Castagnino and Lombardi, solutions of the type (\ref{expcon}) are temporally asymmetric, since there is no cosmic-time hypersurface that splits spacetime into two halves, one the mirror image of the other\footnote{This is clearly possible in (\ref{cycloid}) if one takes the hypersurface $t = t_{\text{max}}$.}. Therefore, they argue, in these spacetimes it is possible to distinguish between past and future exclusively on the basis of properties intrinsic to the manifold itself (like its geometrical and topological structure) and with no reference to thermodynamics or any other nontemporal feature. In particular, either of the two solutions in (\ref{expcon}) would exhibit an arrow of time by the sole fact of being temporally asymmetric. 

We don't believe this to be the case. Temporal asymmetry entails a difference between the two orientations of time, but still does not give a general method to select one. Indeed, in an \textit{exactly} homogeneous and isotropic FRW spacetime, like the ones in (\ref{expcon}) and (\ref{cycloid}), the Weyl curvature tensor is identically zero at all times, meaning that the entropy stays (roughly) constant. Even if this spacetime is temporally asymmetric, we would not expect it to exhibit an arrow of time, because it fails to meet the requirement of increasing entropy. On the other hand, a perturbed FRW solution with small initial inhomogeneities, besides being temporally asymmetric, is also \textit{entropically} asymmetric, a fact that can be characterized geometrically by the monotonic increase of the Weyl contribution in one direction of cosmic time. It is this last feature that gives rise to the arrow of time: the spacetime geometry merely provides the right setting upon which thermodynamics can act. The FRW example we have provided is obviously an idealized case, but it goes to show that temporal asymmetry is a necessary but not sufficient condition for the existence of an arrow of time on the spacetime manifold. 

The position we take is that spacetime needs to obey certain geometrical and topological conditions so that a thermodynamic arrow of time may be defined on it, but it is unequivocally the thermodynamic arrow that determines the direction of time. The sole fact that the spacetime manifold is temporally asymmetric does not provide one with a mean of selecting a particular orientation of time (\S \ref{past}): in the expanding universe of (\ref{expcon}) time could equally well flow from the singularity to infinity or from infinity to the singularity and there is really no way of choosing which one is the arrow an observer will experience in this universe on the basis of geometry alone. One can only do this by invoking a \textit{physical principle}, like the second law of thermodynamics. Were the Big Bang singularity not smooth, as required by (\ref{PHW}), time in our universe would flow in the opposite direction, or not at all. Geometrically, as well as physically, there is really no way of distinguishing the two solutions in (\ref{expcon}) as regards the arrow of time. One can only do so by looking more closely at the inhomogeneities that give rise to different entropy gradients. 

In \S \ref{past} we argued that the entropy requirement can be reformulated as a geometric condition: the Weyl curvature hypothesis. This does not contradict the idea that the arrow of time arises from the second law. Rather, it reflects the fact that the entropy of the gravitational field is inherently geometric, since gravity itself is described by a geometric theory. Time is fundamentally tied to the large-scale geometry and topology of the universe, which must allow for a global asymmetry in the cosmic time direction. However, it is not just \textit{any} geometric asymmetry of spacetime that will yield an arrow of time, but only the particular asymmetry encoded in one specific geometric object (the Weyl curvature tensor) which has a direct link to the entropy of the gravitational field.

There is another reason to believe that entropy must enter somehow into the definition of the arrow of time and this has to do with memory \cite{Wolpert, Rovelli2}: the formation, stability, and accessibility of memories, whether in physical records, biological systems, or cognitive processes, require irreversible dynamics driven by an entropy gradient. The possibility of encoding and retrieving information about past states depends on the dissipation of free energy into heat, so that in a perfectly reversible world, no memory could persist. This ties the thermodynamic arrow directly to the psychological arrow of time: our perception of temporal flow, grounded in the accumulation of records of the past but not the future, emerges from the same entropic asymmetry that shapes the universe at large. Without increasing entropy, neither physical traces nor subjective experience of time’s direction could exist\footnote{Castagnino and Lombardi argue instead that, once a global temporal asymmetry is found, the past-future distinction is conventional and can be expressed locally as a four-dimensional energy flow, without invoking entropy. We agree this is a coherent view, but maintain that entropy is essential to connect global asymmetry with irreversibility, memory, and the experienced passage of time.}.

\section{The Problem of Time}\label{POT}

The simple classical picture presented in the previous sections is complicated, as it often happens, by quantum mechanics\footnote{We assume in this section that gravity is indeed quantized.}. We have shown previously that the passage of time, as commonly understood, is a feature of particular class of (cosmological) solutions of Einstein's equations, and, as such, it is dependent on the geometrical and topological properties of spacetime as a whole. In quantum gravity, however, the universe is really described by a wave function \cite{HartleHawking}, which encodes a superposition of different geometries (and topologies). Of course, we expect the universe to be classical at all but the very early stages of its evolution; nonetheless, the question of whether it is meaningful to talk about a flow of time near the initial singularity, where the quantumness of spacetime is most visible, is a significant one.

The class of solutions we consider are globally hyperbolic universes with topology $\mathbb{R}\times\Sigma$ where $\Sigma$ is the constant-cosmic-time hypersurface. The most general metric decomposition for a globally hyperbolic spacetime $\mathcal{M}$ with topology $\mathbb{R} \times \Sigma$ is \cite{Jha, Carlip}
\begin{equation}\label{canonical}
    ds^2 = (-N^2 + N^i N_i) d\tau^2 +2N_i d\tau dx^i + h_{ij} dx^i dx^j,
\end{equation}
where $N$ and $N^i$ are the lapse function and shift vector, respectively. The lapse function measures the difference between coordinate time and proper time on curves normal to $\Sigma$, while the shift vector measures how spatial coordinates shift from one slice to the other. Note that choosing $N^i = 0$ gives the synchronous spacetime of (\ref{cosmicmetric}).

In the canonical decomposition (\ref{canonical}), the Einstein-Hilbert action becomes
\begin{equation}
    S_{EH} = \frac{1}{16 \pi G} \int d\tau d^3x N \sqrt{h} \left( R^{(3)} + K^{ij}K_{ij} - K^2 \right),
\end{equation}
where $R^{(3)}$ is the 3D Ricci scalar of the spatial slice, $K_{ij}$ the extrinsic curvature, and $K \equiv h^{ij} K_{ij}$ the trace of the extrinsic curvature. 

The Lagrangian density then is
\begin{equation}\label{lagrangian}
    \mathcal{L}_{EH} = \frac{\sqrt{h} N}{16 \pi G} \left( R^{(3)} + K^{ij}K_{ij} - K^2 \right),
\end{equation}
with
\begin{equation}\label{extrinsic}
    K_{ij} = \frac{1}{2N} \left( \dot{h}_{ij}- D_i N_j - D_jN_i \right),
\end{equation}
where $D_i$ is the covariant derivative with respect to the metric $h_{ij}$.

The canonical momentum conjugate to the spatial metric is defined as
\begin{equation}\label{canonicalp}
    \pi^{ij} \equiv \frac{\delta \mathcal{L}}{\delta \dot{h}_{ij}} = \frac{\sqrt{h}}{16 \pi G} \left( K^{ij} - h^{ij} K\right),
\end{equation}
while the Hamiltonian density is defined as the Legendre transform
\begin{equation}\label{hamiltonian}
    \mathcal{H} =   \pi^{ij} \dot{h}_{ij} - \mathcal{L}.
\end{equation}
Inserting (\ref{lagrangian}), (\ref{extrinsic}) and (\ref{canonicalp}) into (\ref{hamiltonian}) gives
\begin{equation}
    \mathcal{H} = N \mathcal{C} + N_i \mathcal{C}^i,
\end{equation}
where
\begin{equation}\label{hamiltonianconstraint}
    \mathcal{C} = \frac{16 \pi G}{\sqrt{h}}\left( \pi^{ij} \pi_{ij} - \frac{1}{2} \pi^2\right) - \frac{\sqrt{h}}{16 \pi G} R^{(3)}
\end{equation}
is the Hamiltonian constraint and
\begin{equation}\label{pconstraint}
    \mathcal{C}^i = - 2 D_j \pi^{ij}
\end{equation}
the momentum constraint. One can rewrite (\ref{hamiltonianconstraint}) as
\begin{equation}
    \mathcal{C} = 2 \kappa \, G_{ijkl} \pi^{ij} \pi^{kl} - \frac{\sqrt{h}}{2 \kappa} R^{(3)},
\end{equation}
where $\kappa \equiv 8 \pi G$, and
\begin{equation}
    G_{ijkl} = \frac{1}{2 \sqrt{h}} \left( h_{ik} h_{jl} + h_{il} h_{jk} - h_{ij} h_{kl}\right)
\end{equation}
is known as the Wheeler–DeWitt metric. This follows from the fact that 
\begin{equation}
    G_{ijkl} \pi^{ij} \pi^{kl} = \frac{1}{\sqrt{h}} \left( \pi^{ij} \pi_{ij} - \frac{1}{2} \pi^2 \right).
\end{equation}
Variation of the Hamiltonian density with respect to $N$ and $N^i$ yields the constraint equations
\begin{equation}\label{constraints}
    \mathcal{C}=0, \quad \mathcal{C}^i = 0.
\end{equation}
In order to quantize equations (\ref{constraints}) we promote $\pi^{ij}$ to the functional derivative operator
\begin{equation}
    \pi^{ij} \rightarrow - i \frac{\partial}{\partial h_{ij}}.
\end{equation}
Demanding that the operator version of $H$ annihilate the wave function $\Psi(h_{ij})$ gives the Wheeler-DeWitt equation in its canonical form\footnote{We ignore here factor-ordering issues.}
\begin{equation}\label{WDW}
    \left(-2\kappa G_{ijkl} \frac{\partial^2}{\partial h_{ij} \partial h_{kl}} - \frac{\sqrt{h}}{2 \kappa} R^{(3)} \right) \Psi(h_{ij}) = 0,
\end{equation}
or, more concisely, $\hat{H} \Psi = 0$, where $\Psi$ is the wave function of the universe. 

Equation (\ref{WDW}) is valid in vacuo. In the presence of matter, there is an additional contribution due to matter fields of the form
\begin{equation}
    S_m = \int d\tau d^3x \left[ \pi_\phi \dot{\phi} - N \left(\frac{\pi_\phi^2}{2 \sqrt{h}} + \frac{\sqrt{h}}{2} h^{ij} \partial_i \phi \partial_j \phi + \sqrt{h} V(\phi)\right) - N^i \pi_\phi \partial_i \phi\right],
\end{equation}
where $\pi_\phi$ is the canonical momentum of the matter field, and $V(\phi)$ its potential. The Wheeler-DeWitt equation then becomes
\begin{equation}\label{WDW2}
    \left(-2\kappa G_{ijkl} \frac{\partial^2}{\partial h_{ij} \partial h_{kl}} - \frac{\sqrt{h}}{2 \kappa} R^{(3)} - \frac{1}{2 \sqrt{h}} \partial_\phi^2  +\frac{\sqrt{h}}{2} h^{ij} \partial_i \phi \partial_j \phi+ \sqrt{h} V(\phi) \right) \Psi(h_{ij}, \phi) = 0.
\end{equation}
There is no explicit time variable in this equation. The wave function encodes correlations among geometric and matter degrees of freedom without ever truly evolving. This is known as \textit{the problem of time} \cite{Kuchar, Isham, Rovelli}:
\begin{quote}
\textbf{Problem of Time}\\
How do we recover temporal dynamics from a fundamental equation that knows none?
\end{quote}
In a way, this is nothing new. General Relativity is reparametrization invariant, as explained in \S \ref{Weyl}, meaning one can relabel coordinates arbitrarily and physics remains unaffected. Hence, coordinate time is not physically meaningful by itself. One has to rely instead on thermodynamic time in order to retrieve a coherent notion of temporal succession and irreversibility (future different from the past), as we saw in \S \ref{past}. 

\begin{figure}
  \centering
  \includegraphics[width=0.9\textwidth]{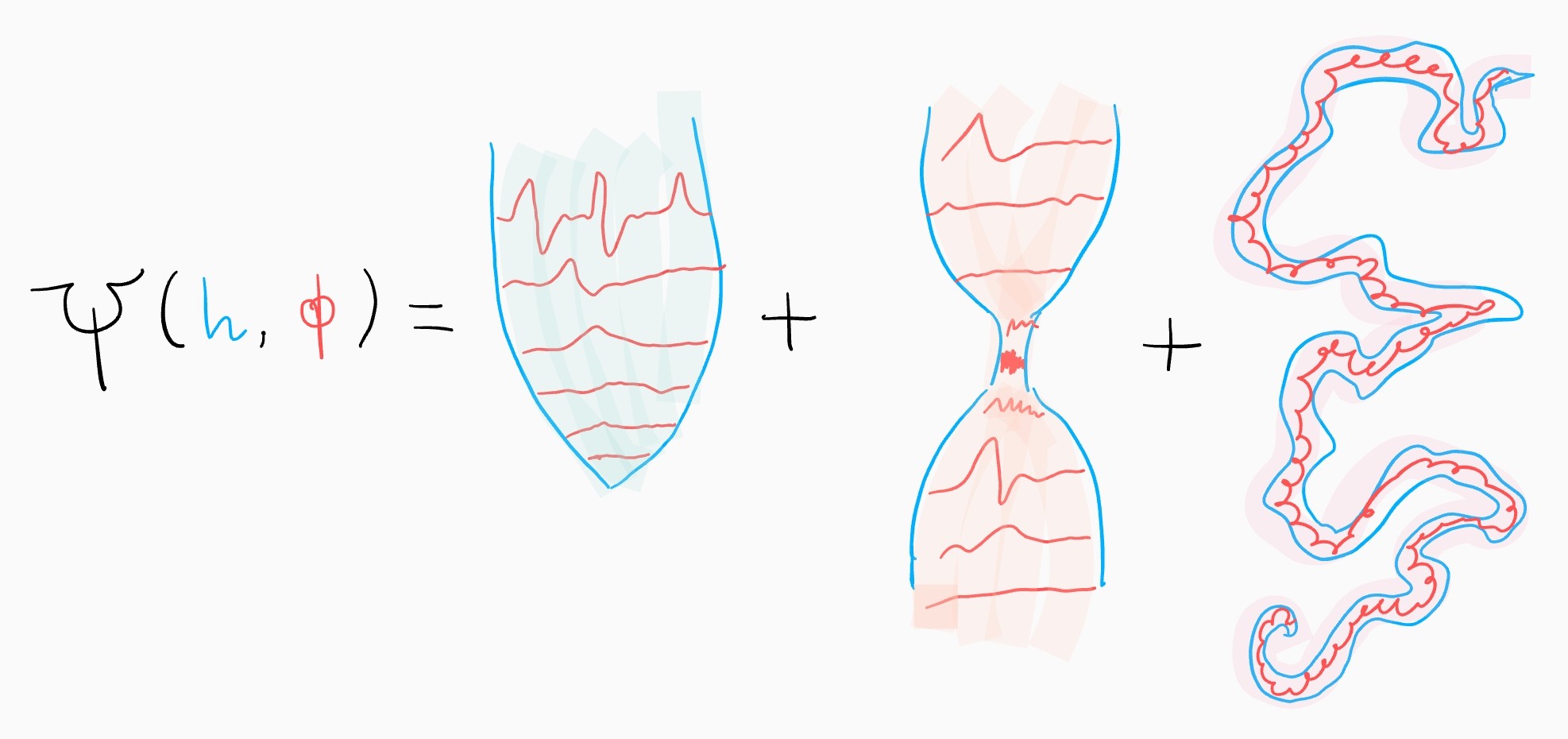}
  \caption*{In quantum gravity we can no longer talk about a single geometry for the universe. We are forced to deal with a universal wave function, which describes a superposition of different geometries (and topologies). The problem of time is therefore doubled, in a way, since we both need to explain how a particular geometry for the universe emerges out of the primordial quantum indeterminacy, \textit{and} how time itself emerges in that specific geometry, which is the ``classical'' problem.}
  \label{fig:quantum}
\end{figure}

However, the procedure that made it possible for us to pick a preferred foliation and with it a notion of cosmic time that every observer can agree upon, which was the prerequisite to define thermodynamic time consistently by singling out the direction of increasing entropy, is not that easily transferred to quantum gravity, as we will now see.

Take again the FRW metric\footnote{Note that this is just (\ref{FRW+}) with the addition of the lapse function $N$.}
\begin{equation}
    ds^2 = -N(\tau)^2 d\tau^2+ a^2(\tau) \gamma_{ij} dx^i dx^j,
\end{equation}
with $h_{ij} = a^2(\tau) \gamma_{ij}$. The metric is clearly invariant under time-reparametrizations $\tau \rightarrow f(\tau)$, as one can just absorb the additional factor into a redefinition of the lapse function. Given that the shift vector $N^i = 0$, the extrinsic curvature (\ref{extrinsic}) is
\begin{equation}
    K_{ij} = \frac{1}{2N} \dot{h}_{ij} = \frac{a \dot{a}}{N} \gamma_{ij}, \quad K = \frac{3 \dot{a}}{N a}.
\end{equation}
Plugging this into (\ref{canonicalp}) gives
\begin{equation}
    \pi^{ij} = \frac{-\sqrt{h}\dot{a}}{N \kappa a} h^{ij}, \quad \pi=-\frac{3 \sqrt{h} \dot{a}}{N \kappa a}.
\end{equation}
Thus,
\begin{equation}
    \pi^{ij}\pi_{ij} - \frac{1}{2} \pi^2 = -\frac{3}{2} \left( \frac{\sqrt{h} \dot{a}}{N \kappa a}\right)^2.
\end{equation}
Now, given that $R^{(3)}=6k/a^2$, (\ref{constraints}) gives\footnote{Note that the momentum constraint $\mathcal{C}^i = 0$ is automatically satisfied.}
\begin{equation}\label{hamiltoneq}
    \left(\frac{\dot{a}}{Na}\right)^2 = \frac{\kappa \rho}{3} - \frac{k}{a^2},
\end{equation}
where $\rho$ is the energy density of the cosmic fluid. Classically, we can just choose the cosmic-time gauge $N(\tau)=1$, which is tantamount to selecting $\tau = t$, where $t$ is the cosmic time, i.e. the proper time of comoving observers. Then, (\ref{hamiltoneq}) reduces to the usual Friedmann equation.

In the context of quantum mechanics, things are not so simple. By virtue of the highly symmetric FRW ansatz, the Wheeler-DeWitt equation (\ref{WDW2}) reduces to\footnote{Technically, the minisuperspace approximation, as it is known, assumes that the infinite-dimensional phase space of (\ref{WDW2}) can be reduced to a finite-dimensional one by considering only the largest wavelength modes, i.e. only those modes that are comparable to the size of the universe. In a cosmological setting, this is a good approximation.}\cite{Tkach}
\begin{equation}\label{timeless}
   \left(\frac{\kappa}{12 a} \partial_a^2 - \frac{1}{2a^3}\partial_\phi^2 -\frac{3ka}{\kappa} + a^3 V(\phi)\right) \Psi(a, \phi) = 0.
\end{equation}
Note that, contrary to the classical case, we cannot gauge-fix $N=1$ here, as the lapse function $N$ is a Lagrange multiplier. There is no remaining gauge freedom at the quantum level; the wave function is frozen in time.

One possible solution to the problem is through a process known as \textit{deparametrization}. Even though (\ref{timeless}) is technically timeless, time can be reintroduced relationally: the scale factor $a$ can play the role of an internal clock, providing a monotonic evolution parameter. Instead of $\Psi(a,\phi)$, one can reinterpret the state as evolving conditioned on $a$:
\begin{equation}
    \Psi(a,\phi) \rightarrow \Psi(\phi | a),
\end{equation}
and treat the scale factor as an emergent internal clock.

This can be done by introducing a semiclassical ansatz for the full wavefunction $\Psi(a,\phi)$ of the form \cite{Halliwell} 
\begin{equation}\label{ansatz}
    \Psi(a, \phi) = e^{i S_0(a)/\hbar} \psi(a, \phi),
\end{equation}
where $S_0(a)$ is the classical action satisfying the Hamilton-Jacobi equation for gravity, and $\psi(a, \phi)$ the wave function encoding quantum matter fields propagating on the classical gravitational background. The decomposition allows us to perform a WKB expansion in Planck's constant.

Substituting (\ref{ansatz}) into (\ref{timeless}), we get an expansion in $\hbar$\footnote{Remember the implicit factors of $\hbar^2$ in the kinetic terms of (\ref{timeless}). Also, it is assumed that $\tilde{V}(\phi) \equiv V(\phi)-V(\phi_c)$.}. Collecting the $\hbar^0$ terms we obtain \cite{Kiefer}
\begin{equation}\label{HJ}
    -\frac{\kappa}{12 a} S_0'^2 - \frac{3ka}{\kappa} + a^3 V(\phi_c) = 0,
\end{equation}
where $\phi_c$ denotes the homogeneous scalar value along the classical trajectory. Equation (\ref{HJ}) is the classical Hamilton-Jacobi equation for the scale factor.

At order $\hbar^1$, and neglecting second derivatives of slowly varying terms with respect to the scale factor, we get the equation
\begin{equation}\label{HJdecomp}
    \frac{i \hbar \kappa}{6 a} S_0' \partial_a \psi - \frac{\hbar^2}{2a^3} \partial_\phi^2 \psi + a^3 \tilde{V} \psi = 0.
\end{equation}
Then, defining the semiclassical time $t$ by
\begin{equation}
    \partial_t \equiv -\frac{\kappa}{6a} S_0' \partial_a,
\end{equation}
we can write (\ref{HJdecomp}) as
\begin{equation}
    i \hbar \partial_t \psi = \left( -\frac{\hbar^2}{2a^2} \partial_\phi^2 + a^3 \tilde{V}(\phi)\right) \psi,    
\end{equation}
the usual Schrodinger equation which describes the quantum evolution of the matter fields $\phi$ on the classical background $a(t)$ via the wave-function(al) $\psi(t,\phi)$.

Deparametrization relies on the spacetime geometry to provide a notion of time. This is similar to what we did classically in \S \ref{Weyl} where we defined cosmic time as the synchronous time of comoving observers. In a way, the semiclassical approach to the problem of time simply reproduces the classical Weyl's solution in a quantum setting, and, as such, is not entirely satisfactory. For one thing, given the nature of the WKB and minisuperspace approximations, this is a useful approach in non-primordial large-scale cosmologies only \cite{Anderson}, so far from the strong quantum gravity regime. The semiclassical solution serves to show that a consistent notion of cosmic time can be defined in a quantum scenario \textit{if} spacetime is already assumed to be large and classical, and merely disturbed by quantum fluctuations in its gravity and matter distributions.

The main question, namely that of how the particular geometry and topology of the universe emerged from the quantum foam, goes unanswered. If there is indeed a connection between the large scale structure of the universe and the flow of time, quantum gravity likely holds the key as to how its unique shape came about, a fact that in classical physics would otherwise be relegated to the unimaginative territory of ``initial conditions''.

\section{Conclusions}\label{conclusions}

We have argued that time as commonly understood is deeply related to the large-scale structure of the universe.

Time in General Relativity is a mere coordinate that labels different spacetime events. Different observers will label events differently, and will have a different notion of time, for example a different plane of simultaneity. There is nothing in the theory that allows one to pick a (universal) time direction in the general case, let alone distinguish between past and future. 

The arrow of time is instead an emergent phenomenon set by the second law of thermodynamics. However, defining a consistent arrow of time on a generic spacetime manifold is only possible under very special circumstances, namely whenever the universe obeys certain geometrical and topological conditions:
\begin{itemize}
    \item The existence of a preferred plane of simultaneity requires the vorticity tensor to vanish everywhere.
    \item The existence of a fundamental difference between past and future, i.e. time asymmetry, requires the universe to be geodesically incomplete with a regular initial singularity.
    \item For the two conditions above to hold \textit{globally} on the whole manifold, the universe needs to be time-orientable and chronological.
\end{itemize}
Whenever the universe obeys these requirements, the arrow of time can be identified with the direction of increasing entropy, which emanates from the regular initial singularity (the Big Bang) and follows the expansion of the universe. If the universe is also topologically trivial, the arrow is globally well-defined and chronology is protected. 

According to this view, what we understand as time (change, flux, becoming, etc.) is not tied in any way to fundamental physics, but is instead a product of the distribution of matter and energy in the whole universe, or, equivalently, since matter and energy determines geometry, of its shape. A simple example is given by irreversibility, which is one of the basic features of our experience of time, but is clearly not a fundamental physical law; it can only be retrieved as an emergent, statistical property of macroscopic systems. As the largest conceivable macroscopic system, the universe itself becomes the natural context for understanding temporal phenomena. Our conjecture adopts the philosophical stance that the passage of time is not a feature of microphysics, but rather a cosmological effect, emerging from and intrinsically connected to the local and global structure of spacetime.

While classically this is enough to give a reasonable instantiation of the A-theory ontology of time, quantum mechanics complicates matters. The fundamental equation of quantum cosmology, the Wheeler–DeWitt equation, is notoriously timeless, and reintroducing time in the description of a quantum system is an infamously difficult problem, more so than its classical counterpart. Many approaches to solving the ``problem of time'' have been tried in the past, each with their strengths and weaknesses. The one that is most akin to the perspective of this paper is the \textit{Tempus Post Quantum} paradigm of \cite{Anderson}, in which time is not present at the fundamental level, but can nevertheless emerge in the quantum regime. Specifically, we presented the semiclassical approach to the problem, in which time is set by the expansion of the universe and is thus tied to geometry. While this approach seemingly works, it does so only under the assumption that the universe is already large and classical, and thus cannot be taken as a full resolution of the problem. A full quantum gravity solution would explain how and why the precise geometrical and topological structure compatible with the existence of time emerges in the first place.

Quantum gravity aside, a natural question one might ask is whether there could be other natural phenomena that are fundamentally tied to the shape of the universe. We plan to come back to this in future work.

\section*{Acknowledgment}
I am grateful to Olimpia Lombardi for introducing me to her work on the global arrow of time, and for the enlightening discussions we shared.

\end{document}